# Control of magnetism in cobalt nanoparticles by oxygen passivation

D. Srikala[1], V. N. Singh[2], A. Banerjee[3], B. R. Mehta[2], and S. Patnaik[1]

[1] School of Physical Sciences, Jawaharlal Nehru University, New Delhi 110067, India

[2] Thin Film Laboratory, Department of Physics, Indian Institute of Technology Delhi, New Delhi 110016, India

[3] UGC-DAE Consortium for Scientific Research, University Campus, Khandwa Road, Indore 452017, India

## Abstract

We report on the preparation of ferromagnetic cobalt nanospheres with antiferromagnetic oxide capping layer and its implication for the variation in magnetic property. The *hcp* cobalt nanospheres were prepared by thermal decomposition of cobalt carbonyl in the presence of organic surfactants. The spherical nanoparticles thus prepared were oxidized to grow antiferromagnetic layers of varying composition and thickness on top of cobalt spheres. High resolution transmission electron microscopy confirmed growth of $Co_3O_4$ in one case and CoO in another case. Strong exchange anisotropy and enhanced coercive field was observed due to the core-shell structure in Co-CoO system. On the other hand only a marginal improvement was seen in Co-$Co_3O_4$ system. A low temperature paramagnetic behavior was also observed that is interpreted in the framework of crystal defects in the oxide shell.

**Introduction**

The potential application of magnetic nanoparticles (NPs) in high density storage devices and medicine is rather restrained today because of issues with its temporal stability.[1,2] The instability occurs because the anisotropic energy per particle, responsible for maintaining the direction of magnetic moment in the presence of thermal fluctuations, decreases rapidly with particle size. This leads to stray flipping of the moments and the onset of *superparamagnetic* behavior at temperatures far below the ferromagnetic to paramagnetic transition temperature ($T_C$). For example, bulk cobalt has a Curie temperature ($T_C$) of 1388 K where as for 4 nm cobalt nanoparticles the temperature corresponding to the loss of ferromagnetic behavior is reported to be ~ 10 K.[3] On the other hand, the concept of exchange bias or enhanced unidirectional magnetocrystalline anisotropy has evolved over the last 50 years and today it finds prevalent application in information storage industry.[4] It refers to the manifestation of an exchange coupling across a ferromagnet – antiferromagnet (FM - AFM) interface. The experimental reflection of this is an asymmetric M - H loop when the sample is field cooled through the Néel temperature of the antiferromagnet vis á vis a symmetric loop in zero field cooled state. The origin of this enhanced anisotropic energy is the torque exerted by the local uncompensated AFM spins onto the FM spins at the interface.[5-7] This net unequal number of up and down spins at the interface depends on anisotropy, spin configuration, and roughness of the material and the resulting exchange coupling between FM nanoparticles and AFM shells provide a source for additional anisotropy. This leads to sustenance of ferromagnetic behavior up to much higher temperature in the presence of thermal fluctuations. In this paper, we compare this enhancement of unidirectional magnetocrystalline anisotropy of Co nanoparticles with different oxide layers (CoO and $Co_3O_4$) synthesized through oxygen passivation. Both CoO and $Co_3O_4$ are antiferromagnets with Néel temperature 293 K and 30 K respectively. Our results show that at T = 5 K, a shell of CoO leads to significant increase in exchange bias as compared to $Co_3O_4$. The blocking temperature for 14.5 nm particle is found to be ~ 150 K. An enormous increase in coercive field, the critical parameter for magnetic storage media, is also observed.

**Experiments**

The spherical cobalt nanoparticles used in this study were prepared by rapid pyrolysis of an organometallic precursor in the presence of organic surfactants. Besides tailoring the size of the particles, it has recently become possible to control the shape of the magnetic nanoparticles using this technique.[8-12] Samples of varying size and surfactant layers were prepared by optimizing two reaction parameters such as the surfactant to precursor volumetric ratio and the reaction time. Subsequently the samples were sonicated for 20 minutes to remove the surfactant capping layer. This was followed by exposure to atmospheric oxygen to grow an antiferromagnetic shell on top of Co nanospheres. Here we concentrate on two samples that show varying degree of exchange bias depending on the oxide layer on the particles. The passivated samples are a) Co core with $Co_3O_4$ shell (POCN1) and, b) Co core with CoO shell (POCN2). The formation of different oxide layers is facilitated by selective growth of crystal faces in the presence of differing concentrations of surfactants. For the first set of cobalt nanoparticles (POCN1), the starting solution contained two surfactants, e.g. 0.2 g of trioctylphosphinoxide (TOPO, 99%) and 0.1 ml of oleic acid that were mixed in 12 ml of 1,3-dichlorobenzene (DCB) and heated at 180°C. Then 0.4 g of $Co_2(CO)_8$ (octacarbonyl dicobalt, stabilized with 1-5 % hexane) was dissolved in 3.6 ml of DCB and rapidly injected into the refluxing bath. The decomposition and nucleation occurred rapidly upon injection. The solution was heated for 15 min and then cooled down to room temperature. Here the surfactant mixture modulates the relative growth rates across the crystal axes in order to yield a specific shape of nanoparticle. Moreover, oleic acid with its carboxylic group adsorbs tightly onto the surface of metallic particles and impedes the particles to grow in size, whereas TOPO acts as selective absorber, altering the relative growth rates across different faces of the crystals. The ratio of strongly and weakly binding surfactants is adjusted to control NP size. For the second specimen (POCN2) a solution containing 0.3 g of TOPO and 0.1 ml of oleic acid in 12 ml of DCB was heated to 180 °C and subsequently 0.4 g of $Co_2(CO)_8$ dissolved in 5 ml of DCB was injected into the refluxing bath. The reaction solution was heated for 25 min and then cooled to room temperature. It is to be noted that the surfactant to cobalt precursor ratio is fixed at 0.7 and 0.9 for POCN1 and POCN2 respectively. Both the reactions were carried out under a continuous flow of high purity argon gas. The size and shape of Co nanoparticles were studied by high resolution transmission electron microscopy using a technaiG$^2$ (200 KV) microscope and crystal structure was identified using electron diffraction spectroscopy (EDS). Magnetization measurements were carried out in a Quantum Design physical property measurement system in the temperature range from 5 to 350 K and in applied magnetic fields up to 5 Tesla. The powder samples for these measurements were obtained from the colloidal black solution by washing several times with equal proportions of ethanol and hexane.

**Results and Discussion**

Fig.1. shows the TEM image of a collection of spherical cobalt nanoparticles with average diameter of 14.5 nm of specimen POCN1. Inset shows the core shell particles with core diameter 8.7 nm, and shell thickness 2.5 nm. HRTEM studies confirm the formation of $Co_3O_4$ shell and the lattice spacing of 1.32 Å is consistent with that of the [711] planes of $Co_3O_4$ (fig. 1b). Each ring of electron diffraction spectroscopy corresponds to a crystal plane as assigned in fig. 1c and thus confirms hexagonal crystal structure of the cobalt core. Similar estimation for POCN2 yielded 10.6 nm and 1.9 nm as core diameter and shell thickness and is shown in fig. 2a. In fig. 2b the crystal plane of the core with lattice spacing 0.193 Å are aligned in [101] direction of *hcp*. The [222] planes correspond to CoO shell. Thus the passivated samples consist of a ferromagnetic *hcp* Co core with antiferromagnetic $Co_3O_4$ and CoO shell in POCN1 and POCN2 respectively. The average diameter of both samples (including the antiferromagnetic layer) is 14.5 nm and we note that particle as a whole is spherical and therefore the shape anisotropy can be neglected.

After establishing the structure of oxidized Co nanoparticles, we next turn to their correlation with exchange bias as ascertained from the magnetic properties. This is derived from the zero-field cool and field cool (ZFC/FC) temperature dependent magnetization curves and from hysteresis loops at different temperatures. The ZFC/FC curves are recorded at 0.5 Tesla and were taken in warming cycle. For FC measurements, the samples were heated up to 350 K and cooled down to 5 K in the presence of the external magnetic field. The ZFC curve in fig. 3 for POCN2 shows a gradual increase of magnetization from 36 K to 136 K as clearly seen in the inset. This suggests a progressive rotation of the magnetic moments of blocked particles towards the field direction as the sample is heated from low temperatures. For both the samples, this behavior is as expected for agglomerated particles and exhibit a broad maximum in ZFC curve.[13] The collective behavior of the system appears only above 36 K, when the thermal energy is sufficiently high to let the magnetic dipoles arrange in a ferromagnetic manner. The divergent point between the FC and ZFC curve in the M vs. T plot is identified as the blocking temperature and we find it to be 122 K and 150 K for POCN1 and POCN2 respectively, which is much higher than 47 K reported for native Co NPs of similar size.[13] A strong paramagnetic contribution is also observed in both the FC and ZFC curves at low temperatures where magnetization increases rapidly. Since there are no magnetic impurities present in the samples as verified by energy dispersive absorption x-ray spectroscopy, this is due to the moments at defect sites present in the nanoparticle surfaces, grain boundaries, and at the interface between the core and the shell. These moments associated with the defects do not relate to the AFM lattice and behave as paramagnetic impurities.[14] The magnetic moment associated with these defects was calculated using Curie-Weiss law analysis by fitting a linear polynomial to the ZFC 1/M vs T data for T < 50 K of fig. 3. The moment per NP of the paramagnetic impurities in the shell is estimated to be 696 $\mu_B$ and 400$\mu_B$ in $Co_3O_4$ and CoO shell respectively. For $Co_3O_4$, the $Co^{+3}$ sites have zero permanent moment and the antiferromegnetism is entirely due to $Co^{+2}$ ions.[15] If each defect is a $Co^{2+}$ ion has a moment of 3.02$\mu_B$[15] and 3.8 $\mu_B$,[14] then the measured moments correspond to about 230 $Co^{2+}$ moments in $Co_3O_4$ and 105 $Co^{2+}$ in CoO shell that are not coupled to the AFM lattice. These defect moments also have an indirect effect on exchange bias according to domain state model which predicts that any deviation from a perfect AFM crystalline structure (away from the interface) can favor the formation of magnetic domains that leads to an increase in exchange bias coupling.[16] A more quantitative analysis on the contribution of these defects to the magnitude of exchange bias observed in our experiments need further investigations.

Fig. 4 shows the ZFC and FC magnetization versus field hysteresis loops of POCN1 and POCN2 at 300 K and at 5 K. As shown in inset of fig. 4a, POCN1 at 5 K exhibits the normal behavior of a symmetrical hysteresis loop when cooled in the absence of a field (ZFC), but when cooled in a strong magnetic field it exhibits a shift along the field axis in the negative direction to the field. This asymmetry, the so called "exchange bias", is estimated to be 132 G in POCN1 (inset Fig. 4a). On the other hand, as seen in inset Fig. 4b, when POCN2 is field cooled, exchange bias increased to 483 Gauss. We note that the magnetic moment in $Co_3O_4$ and CoO arises from $Co^{2+}$ ions at tetrahedral and octahedral lattice sites respectively. The contribution for magnetic moment is from the same ion but the coordination of interfacial spins leads to important changes in the overall magnetic order of the oxygen passivated nanoparticles.

A closer look at the hysteresis loops in fig. 4 also points towards an increased coercivity (half of total width of the loop along the field axis), after the field cooling procedure. This disappears above AFM Néel temperature. High coercivity is an important requirement for the magnetic storage devices. The coercive field $H_c$ at T = 5 K has increased as compared to the coercive fields at 300 K by a factor of 6 in POCN1 and 13 in POCN2. We also observe that both the samples are superparamagnetic at 300 K. The paramagnetic spins of the shells are unable to bias the ferromagnetic spins resulting in null coercivity. These results suggest that the coercivity of the particles is also strongly influenced by the interaction between the Co-oxide shell and the Co core. No saturation magnetization (at T = 5 K) was observed in either sample up to 5 T due to the field dependent susceptibility of antiferromagnetic shell. In fact no saturation would be expected as the AFM layer has large magnetocrystalline anisotropy. At 1.5 Tesla, well above the bare Co saturation, the magnetization is found to be 77.4 emu/g at 5 K and 31.2 emu/g at 300 K for POCN1 and 59.8 emu/g at 5 K and 34.6 emu/g at 300 K for POCN2. This substantial decrease in magnetization as compared to bulk Co ($M_s$ ~ 162 emu/g)[14,17] with decrease in particle size results from various factors that include the increasing NP surface to volume ratio, the presence of antiferromagnetic layer on the surface and to a lesser degree possibly due to chemi-absorption of the surfactant as a carboxylate onto the Co nanoparticle.[18] The decrease in saturation magnetization at higher temperature is indicative of an admixture of ferromagnetic and paramagnetic phase whose relative concentration varies with temperature. Above 2.1 T and 2.5 T, both the ZFC and FC curves merge into a single curve for POCN1 and POCN2 respectively.

The magnetocrystalline anisotropy constant, that describes the preference for spins to align in a particular direction within the particle depends on various anisotropies like shape anisotropy, surface anisotropy, stress anisotropy, and unidirectional anisotropy[19], and is approximately given by $K = 30 k_B T_B / V$, where $T_B$ is the blocking temperature and V is the particle volume. Our results yield $K = 3.16 \times 10^5$ erg/cm$^3$ for POCN1 and $3.89 \times 10^5$ erg/cm$^3$ in POCN2. As the particle size decreases the effective anisotropy energy increases because of smaller core size effects. In our case, in spite of smaller core in POCN1, the effective anisotropy is observed to be smaller when compared to larger core of POCN2. While the enhanced exchange bias in POCN2 appears to be the dominant reason, there are several possible reasons for the reduction of effective anisotropy in POCN1. Since the shape of the particles is spherical, the shape anisotropy will be negligible yet the stress anisotropy, which results from internal stresses due to cooling, may have some contribution to the total value. Moreover, in spherical magnetic nanoparticles the surface contribution related to the radial orientation of the surface spins can be negative resulting in an effective magnetocrystalline anisotropy reduction with decreasing nanoparticle size.[20] The observed coercivity is therefore proportional to the anisotropy constant and inversely proportional to the saturation magnetization. Further, the volume of the Co core in POCN1 and POCN2 is about 21% and 39% of the total volume of the particle but the exchange bias is much larger in POCN2. Thus, more than the volume of the AFM shell, the structure and the interfacial interactions play critical role in controlling the unidirectional magnetocrystalline anisotropy.

In conclusion, spherical cobalt nanoparticles of 14.5 nm diameter were prepared by rapid pyrolysis method. The samples were exposed to atmospheric oxygen to grow antiferromagnetic layers to study the exchange bias effect. HRTEM results confirmed growth of $Co_3O_4$ and CoO in two samples prepared with different surfactant to precursor ratio. The presence of crystal defects in the AFM layers gave rise to low-temperature paramagnetic response in both the samples. Strong exchange bias effects along with high coercivity were observed in the case of CoO capping layer as compared to $Co_3O_4$.

**Acknowledgement:**

DSK acknowledges UGC for financial support. We also acknowledge DST, Government of India for funding the VSM at CSR, Indore and High field magnet facility at JNU, New Delhi. SBP acknowledges support from DST, India under the young scientist scheme. We thank Sulabaha Kulkarni for fruitful discussions.

**References**

(1) S. A. Wolf, D. D. Awscalom, R. A. Buhrman, J. M. Daughton, S. von Molnár, M. L. Roukes, A. Y. Chtchelkanova, and D. M. Treger, Science **2001**, *294*, 1488.

(2) S. D. Bader, Rev. Mod. Phy. **2006**, *78*, 1; C. B. Murray, Shouheng Sun, Hugh Doyle, and T. Betley, MRS Bulletin, December **2001**, *985*.

(3) V. Skumryev, S. Stoyanov, Y. Zhang, G. Hadjipanayis, D. Givord, J. Nogués, Nature **2003**, *423*, 850.

(4) A. E. Berkowitz, K. Takano, J. Magn. Magn. Mater. **1999**, *200*, 552.

(5) J. Nogués, I. K. Schuller, J. Magn. Magn. Mater. **1999**, *192*, 203.

(6) W. H. Meiklejohn and C. P. Bean, Phys. Rev. **1956**, *102*, 1413.

(7) R L Stamps, J. Phys. D: Appl. Phys. **2000**, *33*, R247.

(8) V. F. Puntes, K. M. Krishnan, and A. P. Alivisatos, Science **2001**, *291*, 2115.

(9) C. P. Gräf, R. Birringer, and A. Michels, Phys. Rev. B **2006**, *73*, 212401.

(10) J. Waddell, S. Inderhees, M. C. Aroson and S. B. Dierker, J. Magn. Magn. Mater **2006**, *297*, 54.

(11) Nisha Shukla, Erik B. Svedberg, John Ell, A. J. Roy, Mater. Lett. **2006**, *60*, 1950.

(12) H. Shao, Y. Huang, H. Lee, Y. Suh, C. Kim, J. Appl. Phys. **2006**, *99*, 08N702.

(13) V. F. Puntes, K. M. Krishnan and P. Alivisatos, Appl. Phys. Lett. **2001**, *78*, 2187; V. F. Puntes and K. M. Krishnan, IEEE Trans. Magn. **2001**, *37*, 2210.

(14) J. B. Tracy, D. N. Weiss, D. P. Dinega and M. G. Bawendi, Phys. Rev. B **2005**, *72*, 064404.

(15) W. L. Roth, J. Phys. Chem. Solids **1964**, *25*, 1.

(16) J. Keller, P. Miltényi, B. Beschoten, G. Güntherodt, U. Nowak, and K. D. Usadel, Phys. Rev. B **2002**, *66*, 014431.

(17) J. R. Childress, C. L. Chien, Phys. Rev. B **1991**, *43*, 8089.

(18) N. Wu, L. Fu, M. Su, M. Aslam, K. C. Wong, and V. P. Dravid, Nano Lett. **2004**, *4*, 383.

(19) A. F. Bakuzis, P. C. Morais, J. Magn. Magn. Mater **2001**, *226*, 1924.

(20) A. F. Bakuzis, P. C. Morais, and F. Pelegrini, J. Appl. Phys. **1999**, *85*, 7480.

**Figure Caption**

Fig.1. Bright field TEM micrograph of cobalt nanospheres (POCN1). Inset shows the cobalt nanospheres with a core diameter of 8.7 nm and a shell of $Co_3O_4$ with thickness ~ 2.5 nm, (b) HRTEM image shows the [711] crystal plane of $Co_3O_4$ shell, and (c) Electron diffraction micrograph rings comply with *hcp* Co and each ring is indexed.

Fig.2. (a) TEM image of Co nanospheres with core diameter 10.6 nm and CoO shell of 1.9 nm of POCN2, and (b) HRTEM image shows the [101] crystal planes of Co core and [222] plane of CoO shell

Fig.3. Temperature dependence of zero field cooled and field cooled magnetization of POCN1 and POCN2 at 5 K. A low temperature rise in the magnetization indicates the presence of defects in the AFM component. Inset shows the gradual rotation of magnetic moments towards the field direction from 36 K to 136 K.

Fig.4. Hysteresis loops of POCN1 and POCN2 nanospheres measured at different temperatures. Insets show zero-field cooled and field-cooled (5 kG) magnetization curves at various temperatures exhibiting exchange bias. A substantial exchange bias for the field cooled case is seen in both samples but the magnitude is over 230 % higher in POCN2.

**Fig. 1**

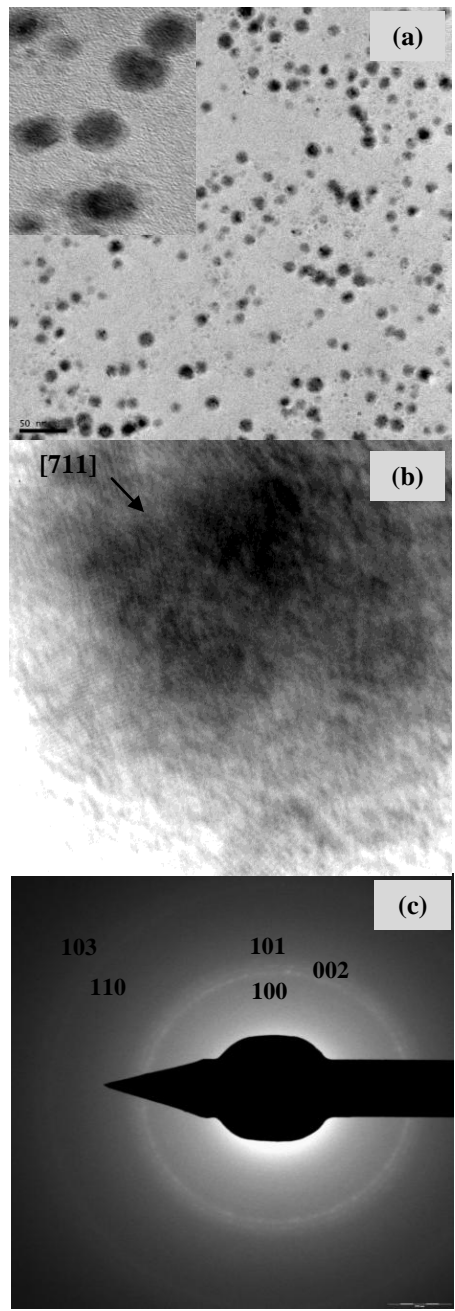

**Fig. 2**

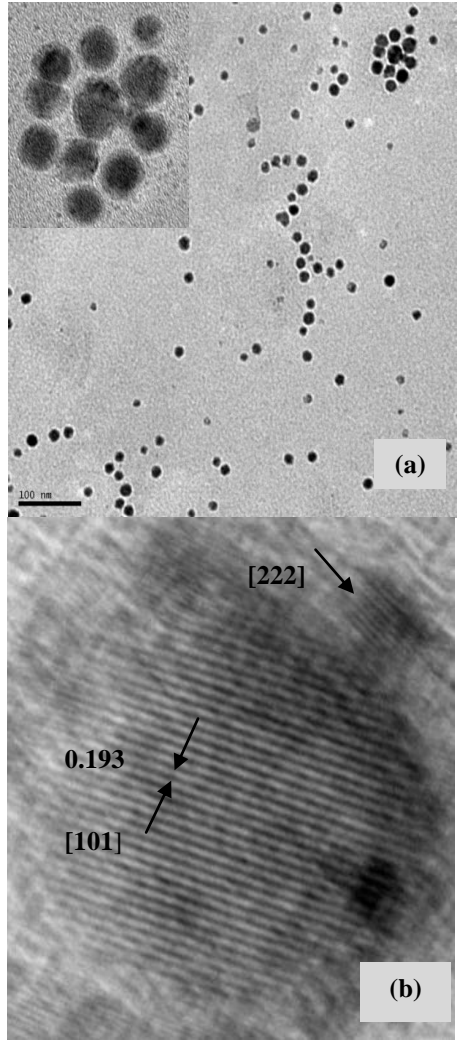

**Fig. 3**

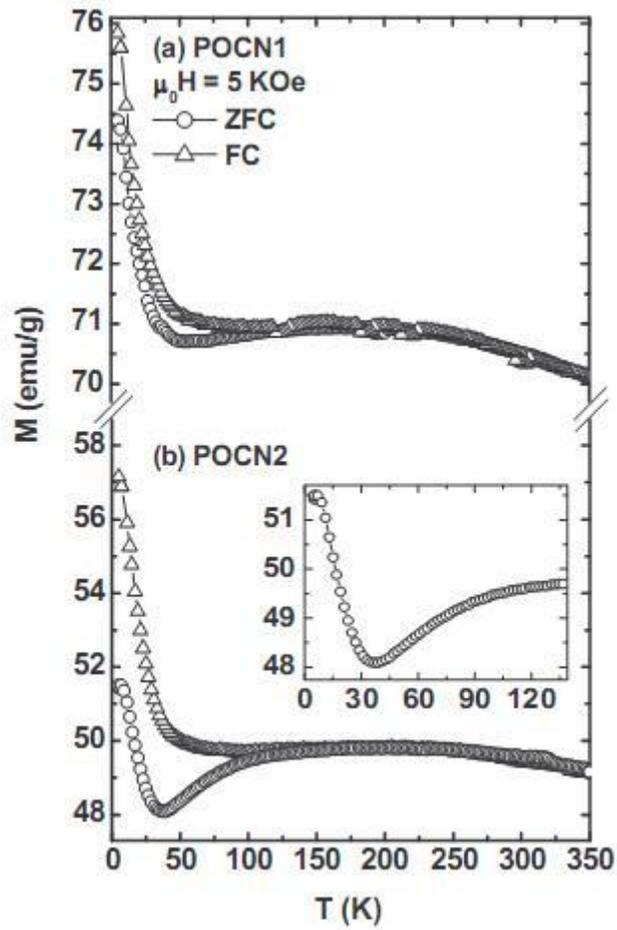

**Fig. 4**

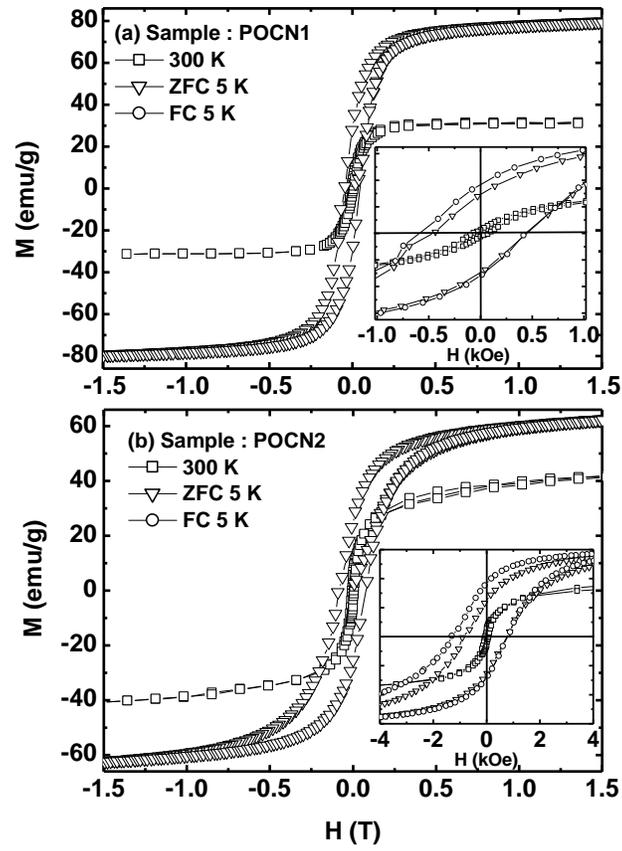